\documentclass[onecolumn,showpacs,floatfix,preprintnumbers,amsmath,amssymb,aps]{revtex4-1}
\usepackage
[
        a4paper,
        left=3cm,
        right=3cm,
        top=3cm,
        bottom=4cm,
]{geometry}
\usepackage{amsmath}
\usepackage{amssymb}
\usepackage{bm}
\usepackage{color} 
\usepackage{graphicx}
\usepackage{pstricks}

\newcommand{\m}[1]{\ensuremath{\left< #1\right>}} 
\newcommand{\lef}{\left(} 
\newcommand{\rig}{\right)} 

\newcommand{\refeq}[1]{eq.\ (\ref{#1})}
\newcommand{\fig}[1]{Fig. \ref{#1}}
\newcommand{\figa}[2]{Fig. \ref{#1}.#2}

\newcommand{\dd}[0]{\mathrm{d}}

\begin{document}

\title{Stochastic Dynamics of Resistive Switching: Fluctuations Lead to Optimal Particle Number}

\author{Paul K. Radtke$^1$, Andrew L. Hazel$^2$, Arthur V. Straube$^3$ and Lutz Schimansky-Geier$^1$} 

\address{$^1$Department of Physics, Humboldt-Universit\"at zu Berlin, Newtonstra\ss e 15, 12489 Berlin, Germany}
\address{$^2$School of Mathematics and Manchester Centre for Nonlinear Dynamics, University of Manchester, Oxford Road, Manchester, M13 9PL, UK}
\address{$^3$Department of Mathematics and Computer Science, Freie Universit\"at Berlin, Arnimalle 6, 14195 Berlin, Germany}

\begin{abstract}

Resistive switching is one of the foremost candidates for building novel types of non-volatile random access memories. Any practical implementation of such a memory cell calls for a strong miniaturization, at which point fluctuations start playing a role that cannot be neglected. A detailed understanding of switching mechanisms and reliability is essential. For this reason, we formulate a particle model based on the stochastic motion of oxygen vacancies. It allows us to investigate fluctuations in the resistance states of a switch with two active zones.
The vacancies' dynamics is governed by a master equation. Upon the application of a voltage pulse, the vacancies travel collectively through the switch. By deriving a generalized Burgers equation we can interpret this collective motion as nonlinear traveling waves, and numerically verify this result.
Further, we define binary logical states by means of the underlying vacancy distributions, and establish a framework of writing and reading such memory element with voltage pulses. Considerations about the discriminability of these operations under fluctuations together with the markedness of the resistive switching effect itself lead to the conclusion, that an intermediate vacancy number is optimal for performance.

\end{abstract}


\maketitle


\section{Introduction}
\begin{figure}
\begin{center}
\includegraphics[scale=.7]{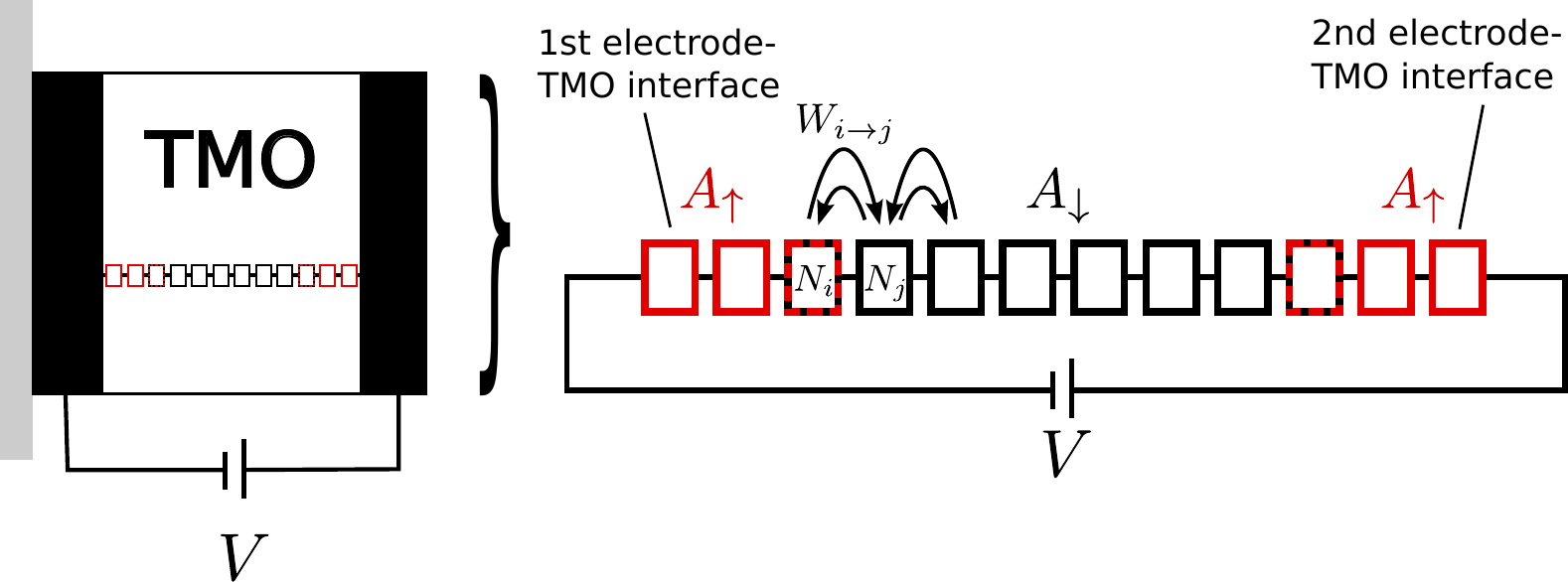}
\caption{Single RS-element consisting of a transition metal oxide electrodes on a substrate material (grey), connected to an external voltage source $V$. The right electrode is defined as having zero voltage. Crossing through the device is a filament of high conduction, whose behavior is responsible for the state of the device. It is modeled as a 1d lattice with three different zones, the bulk (B) which has a small resistance factor $A_{i} = A_\downarrow$, and the interface zones to the electrodes on the left and right, which have large resistance factors $A_i = A_{\uparrow}$, with $A_\downarrow \ll A_\uparrow$. The individual sites are filled with $n_i$ vacancies, that can hop to neighboring positions. }
\label{crossbar}
\label{hopping}
\label{band-structure}
\end{center}
\end{figure}

Resistive switching (\textbf{RS}) refers to a change in the resistance of a dielectric due to the action of an external electric field or an electric flux through the medium. Thereby, the resistance depends on the history of the field or flux passing through the system, hence it can be considered as a hysteretic effect. RS has been observed in a wide range of transition metal oxides (\textbf{TMO}'s), such as manganites $MnO(OH)$ \cite{Chen2005, Ignatiev2006}, perovskites $CaTiO_3$ \cite{Asamitsu1997, Beck2000, Fujii2005} and titanium dioxide $TiO_2$ \cite{Lee2011, Strukov2008MissingMemristor, Ghenzi2013, Yang2009}. Its basic layout is a two terminal device consisting of a TMO sandwiched between two electrodes, as sketched in Fig. \ref{crossbar}. In general the strength of the effect, i.e. the ratio of the high and low resistance, increases with smaller system sizes \cite{Strukov2008MissingMemristor}. 
different realizations are possible. For example, we have phase change memory, consisting of a chalcogenide glass that switches between an amorphous and a crystalline phase \cite{Waser2009, Chua2011}. 

Recently, applications particularly in the semiconductor industry have taken up at a rapid pace. Among the most promising candidates is resistive random access memory (ReRam). Other interesting applications are the integration of logic in memory \cite{Vourkas2016}, enabling  example concepts of neuromorphic computing \cite{Jo2010, Indiveri2013}.  ReRam is expected to provide highly scalable, fast, non-volatile and low cost memory \cite{Waser2009, Jeong2012, Yang2013}. A single such cell is toggled in between its high- and low-resistive state by application of an external voltage or current. Typical implementations for industrial use aim for high density and stack those elements into a 3d nanocrossbar, layered grids of wires with RS cells in between \cite{Vontobel2009, Yang2013, Linn2010}. However, along with high integration and miniaturization, challenges of reliability due to sneak paths and fluctuations become ever more significant. 

To address the problem of read failures and heat dissipation due to sneak paths, 
Linn et. al. \cite{Linn2010} suggested a complementary resistive switch (\textbf{CRS}). Therein, two RS elements are combined anti-serially to one memory cell.  Both its logical states have a high resistance, albeit with differing internal states of the constituent elements. 
Due to the high resistance, the current and associated energy dissipation through the memory cell and the occurence of sneak paths around it are drastically reduced. As such, the concept has  been picked up by various works, see e.g. \cite{Yu2010, Lee2011, Budhathoki2013, Ambrogio2014, Vourkas2014, Radtke2016}. Conceptually, this setup is similar to having a single element with two active switching zones, one at each electrode-TMO boundary, as it will be applied in this work.

Fluctuations can appear externally and internally. They have been studied for phenomenological memristor models
, whose resistance is determined by a single internal scalar variable, which denotes the relative sizes of a high and low and high resistivity area. Internal fluctuations are incorporated by adding white noise to this variable, \cite{Stotland2012, Patterson2016}, with beneficial effects on the RS-effect, such such as increasing the contrast between the resistive states. External fluctuations were studied in the form of noisy impulses switching the states, and depending on the setup can either have a positive  \cite{Patterson2013c, Patterson2015} or detrimental effect \cite{Patterson2016}. As yet, no study of fluctuations has been conducted for a particle based model.

Such an approach is essential to address many characteristics of resistive switching. 
From experimental observations it is known that the functionality of a RS device is determined by the electrode-TMO interface and the distribution of oxygen vacancies \cite{Tsui2004, Szot2006, Nian2007}. In this setting, a one dimensional lattice model in which a probability distribution of the vacancies evolves depending on the external voltage and the local resistance of each lattice site is proportional to its density of vacancies has been proposed in \cite{Rozenberg2010} for manganites, termed the \textit{voltage enhanced oxygen-vacancy migration model} (VEOV-model) by the authors. The Schottky barriers are incorporated by enhancing the vacancies effect near the interfaces compared to the bulk, resulting effectively in a two active switching zones. Further investigation of this model showed shock-wave like behavior, \cite{Tang2014}, made plausible by the formulation of a generalized Burgers equation for the time evolution of the vacancy distribution, which was used to predict the commutation speed of an 
RS element.

of the vacancy distribution.


Hence, our goal is to describe a complementary resistive switch based on this mesoscopic model for discrete particles. 
We elaborate on the requirements to implement such a device, namely as a large contrast between the resistive states, and the reliability against fluctuations. For very small system, the fluctuations will be driven by the inherent stochasticity in the motion of just a few oxygen vacancies. 
This lets us determine a lower limit for the possible level of miniaturization and we can predict an optimal system size for the resistive switch. We proceed as follows:

In section \ref{stochastic_RS} we will formulate the VEOV-model for discrete particles, i.e. the oxygen vacancies. It is governed by a multivariate master equation with nonlinear transition rates. This allows us to consider not only the system dynamics depending on the parameters of the external driving, but also to further examine the fluctuations, whose magnitude is determined by the number of vacancies in the system. 

Subsequently, we investigate the stochastic dynamics of the oxygen vacancies in section \ref{generaldynamics}. Hereby, we pay special attention to the hysteretic effects, which we will quantify by the area of the corresponding hysteresis loops, and on the minimum and maximum resistances the system visits within a period of a periodic driving (cf. section \ref{fluctuations}). Further, we introduce continuous space and derive a nonlinear continuity equation governing the evolution of the oxygen vacancy distribution (cf. section \ref{sectionburgers}). This equation can also be considered a generalized Burgers' equation, and hence the dynamics of the oxygen vacancies are interpreted as nonlinear traveling waves. Also, we successfully numerically integrate said equation to compare it with the results gained in the discrete particle picture, and show how these wave processes affect the electric properties of the device. 

In section \ref{sectionlogic}, we define the logical states of the a resistive switch with two active zones in terms of the underlying particle distributions. This allows us to express the actions of the driving voltage in these terms and develop a framework of how to write and read information in such a system (cf. \ref{sectionreadingop}). The resulting read operation is finally considered for its stability with regard to fluctuations (cf. section \ref{sectionclarity}). Together with the previous results, this leads to an estimate for the optimal particle number.

\section{A stochastic vacancy hopping model for bipolar resistive switching}
\label{stochastic_RS}
We base our considerations on the phenomenological VEOV-model. It focuses on the oxygen vacancy defects together with the influence of the electrode-TMO interface on bipolar switching \cite{Rozenberg2010}. Justified by experimental measurements showing filamentary paths of high conductance \cite{Fujiwara2008}, the memristor is modeled as a one dimensional lattice, i.e. a network of resistors in serial order.

In transition metal oxides the oxygen vacancies affect the resistance in a complex way. On the one hand, they provide dopants and thereby facilitate conductance \cite{Larentis2012}. On the other hand, they disrupt the TM-O-TM bonds. In order to minimize the energetic cost of the vacancies, their place is filled by free electrons from the conduction band, that are localized now in this place and have an energy level below the conduction band \cite{Pan2013}. This resistance effect is more pronounced at the boundary of the TMO, because at a metal-insulator transition the TMO is depleted of freely moving charges, an effect that is also known as Schottky-barrier. Hence taking electrons of the conduction band leads to a higher impact on the local resistance here. 

\begin{figure}
\begin{center}
\includegraphics[scale=0.7]{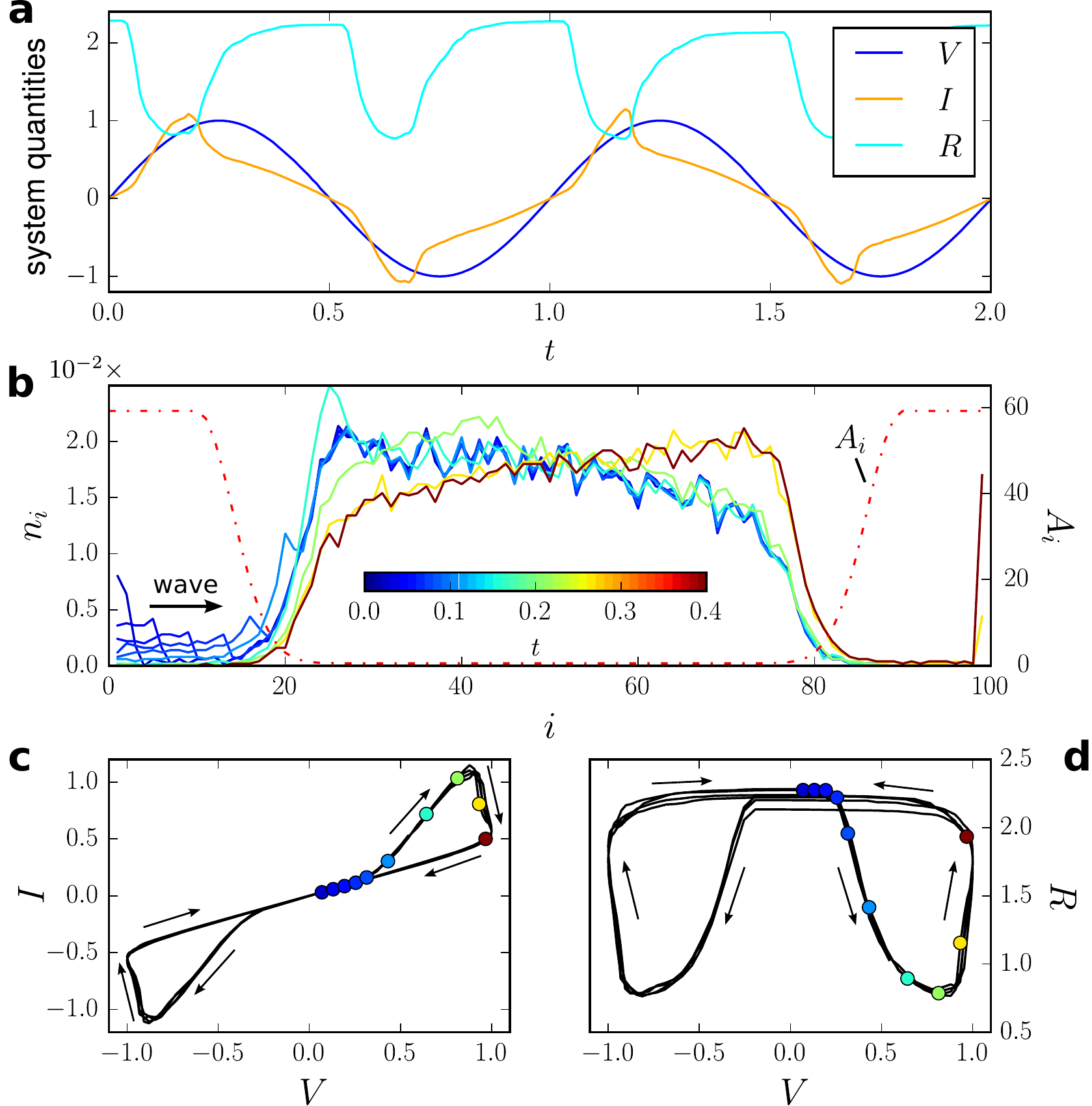}
\caption[Simulation results of resistive switching with sinusoidal voltage driving]{Simulation of stochastic RS-device with sinusoidal voltage driving. Parameters are  $\m{N_i} = 10^{2}$, $N_0=10^4$. (\textbf{a}) Time behavior of the electrical quantities (\textbf{b}) Occupation of the individual lattice sites at different times (indicated by the color scale). Local resistivity $A_i$ is indicated by the red dash-doted line
(\textbf{c}) Hysteresis in the $I-V$ and (\textbf{d}) $R-V$ planes, the colored circles indicate the states for the various occupation distributions of the middle panel, the arrows show the curves' directionality with increasing time.}
\label{dyn-sine}
\end{center}
\end{figure}

For our investigations, the disrupting influence of the vacancies is considered, as it occurs for example in manganites \cite{Chen2005, Ignatiev2006}. We assume that a specific lattice site $i$ has $N_i$ vacancies, every lattice site can hold a total of $N_0$ vacancies (with, in general, $N_i \ll N_0$), and that the resistance of that each site is proportional to its density of vacancies, 
\begin{equation}
n_i := N_i/N_0.
\label{particledensity}
\end{equation}
 The depletion zones at the electrode interface are  taken into account by further multiplication with a resistivity factor $A_i$, where $A_{i}$ in the bulk is small against $A_i$ near the boundaries of the domain. As a result, the local resistance is given by
\begin{equation}
R_{i} = A_{i}\frac{N_{i}}{N_0},
\label{locres}
\end{equation}
from which we gain the total resistance of the lattice by summation over all sites, $R=\sum_i A_i N_i/N_0$. The analytical form of $A_i$ is given by
\begin{equation}
  A(x)=\begin{cases}
    A_\uparrow,  \text{ if } x< x_0 \text{ or } x> L - x_0,\\
    A_\downarrow + (A_\uparrow-A_\downarrow)\lef \frac{1}{2} + \frac{1}{2}\cos\lef2\pi \frac{x-x_0}{L-2x_0}\rig\rig^k\text{ else},\\
  \end{cases}
  \label{A_formula}
\end{equation}
its actual course can be seen in \figa{dyn-sine}{b}, where it is indicated by the dash dotted lines, for a discrete lattice of length $L$, with a length of the depletion zones $x_0$, and with $100$ lattice sites at the positions $i(x)=\mathrm{floor}(100 \cdot x/L)$. Therein, the parameters used are $k=20$, $x_0/L=0.1$ and $A_\uparrow / A_\downarrow = 100$.  

Let  a configuration of lattice vacancies $(N_1, N_2, \cdots, N_L)$ be denoted by $\mathcal{N}$, and by $\{\mathcal{N'}\}$ the set of configurations that arises from $\mathcal{N}$ through a single particle hopping to a neighboring site. This results in the following master equation
\begin{align}
\begin{aligned}
 \frac{\mathrm{d}}{\mathrm{d}t}P(\mathcal{N},t) =  \sum_{\mathcal{\{N'\}}\neq\mathcal{N}}\Big(& -W_{\mathcal{N}\rightarrow \mathcal{N'}} P(\mathcal{N},t)+ W_{\mathcal{N'}\rightarrow \mathcal{N}} P(\mathcal{N'},t)\Big).
\label{master-eq}
\end{aligned}
\end{align}
 An external voltage at the electrodes will induce an electrical current $I$ through the system. It can be used to determine the voltage drop over each lattice site, $\Delta V_{i}$. The oxygen vacancies are quasiparticles holding a charge of $q_{dop}$, whose diffusive motion is biased by the local voltage drop $\Delta V_i$. Agitated, oxygen vacancies now move with a certain probability from one site to another, limited by the number of possible free spaces at their respective target sites. Also, the vacancy requires an activation energy $E_A$ to leave its place in the lattice, that needs to be weighted against its thermal energy $k_B \theta$, with the temperature $\theta$ and the Boltzmann constant $k_B$. Together, we can write down the hopping rate
\begin{equation} 
w_{i\rightarrow i \pm 1}=\Big(1-\frac{N_{i \pm 1}}{N_0}\Big)\exp{\lef\frac{-E_{A} \pm q_\text{dop} \Delta V_{i}}{k_B\theta}\rig}.
\label{transprob}
\end{equation}
The total rate $W_{i\rightarrow j}$ of a particle hop from lattice site $i$ to $j$ is proportional to the number of vacancies at the starting place, $W_{i\rightarrow j} = N_i w_{i\rightarrow j}$. This basic setup is visualized in Fig. (\ref{hopping}).


Without external voltage, the oxygen vacancies will be equidistributed. If the external voltage is changed according to some time protocol, directed motion of the vacancies appears. We will investigate the system for a sinusoidal voltage driving $V(t) = V_0 \sin(2\pi t /T)$ with amplitude $V_0$ and period $T$, the latter setting the scale for all system times. The voltage drop over a lattice site is proportional to its resistance, hence
\begin{equation}
\Delta V_i = V(t)R_i/ R = I(t) R_{i}.
\label{volt-drop-disc}
\end{equation}
Here, the electrical current $I(t)$ is, just as the total resistance $R$, a nonlocal property that depends on the configuration of the entire lattice.

Simulation results for this dynamics are obtained via the Gillespie algorithm \cite{Gillespie1976, Gillespie1977, Feistel1978, Ebeling1979}, which provides a feasible performance since it scales with the transition rates, which themselves are exponentially dependent on the external voltage.

For convenience, we introduce reduced units in which all quantities will be given,
\begin{align}
\hat{V} & = V \cdot  q_\text{dop}/ k_B \theta, \qquad  \hat{R}  = R/R_0 \\
\hat{I}  & = I \cdot  q_\text{dop}R_0/ k_B \theta, \qquad \hat{t} = t/T \\
\hat{E}_A & = E_A / k_B \theta, \qquad\qquad \hat{x} = x/L.
\end{align}
Thereby $R_0 \equiv 1.76 \Omega$ follows by aa resistivity weighted equidistribution of the vacancies, which is explained in Appendix \ref{app-numerics}.  Unless stated otherwise, we choose for the simulations parameters $E_A= 1$, $T = 1 $, $V_0 = 500 $ and $N_0 = 10^4$. Henceforth, all hats are dropped.
 
\section{General Dynamics}
\label{generaldynamics}
The time evolution of a system governed by eqs. (\ref{locres}-\ref{volt-drop-disc}) is depicted in Fig. \ref{dyn-sine}. The electrical current $I$ follows the voltage input, until at a certain threshold value above or below zero the resistance suddenly drops significantly and the current spikes. Hereby, the course of $I$ and $R$ is periodic within each half-cycle of the voltage driving, due to the symmetry of the device.

Looking at the time dependent vacancy occupation distribution \figa{dyn-sine}{b} for the positive half-period of the voltage driving, we see that this corresponds to vacancies moving from the left boundary region of $A_i$ to the bulk. The form of this motion reminds us to a dispersing wave,  as also noted in \cite{Tang2014}, until it reaches the bulk. From there some vacancies finally reach the right boundary region. In the final panels the hysteresis loops in the $I-V$ plane and a two-legged structure in the $R-V$ plane are depicted, cf. \cite{Rozenberg2010}. It reveals that the wave reaching the bulk corresponds to a drop in the device resistance, and a corresponding spike in the current. As soon as the vacancies reach the right side, the resistance increases again. 

As the transport from an accumulation on the sides to the bulk differs in its dynamics from the transport from the bulk to the side, we see the emergence of hysteresis loops in the $I-V$ and $R-V$ diagrams. Also we note that the voltage drop $\Delta V_{i}$ depends on both the external voltage $V$ and the $N_{i}$-configuration of the entire lattice, and as a consequence so does $w_{i\rightarrow j}$. Hence the distribution of the vacancies plays the role of the inherent memory variable of this resistive switching device, much as the boundary positions $w$ does in the HP-memristor \cite{Strukov2008MissingMemristor}. While a transition of oxygen vacancies between lattice sites with different form factors $A_{i}$ exerts a global influence on all reaction rates, jumps between lattice sites with the same $A_{i}$ only affect local rates.

\begin{figure}
\begin{center}
\includegraphics[scale=0.8]{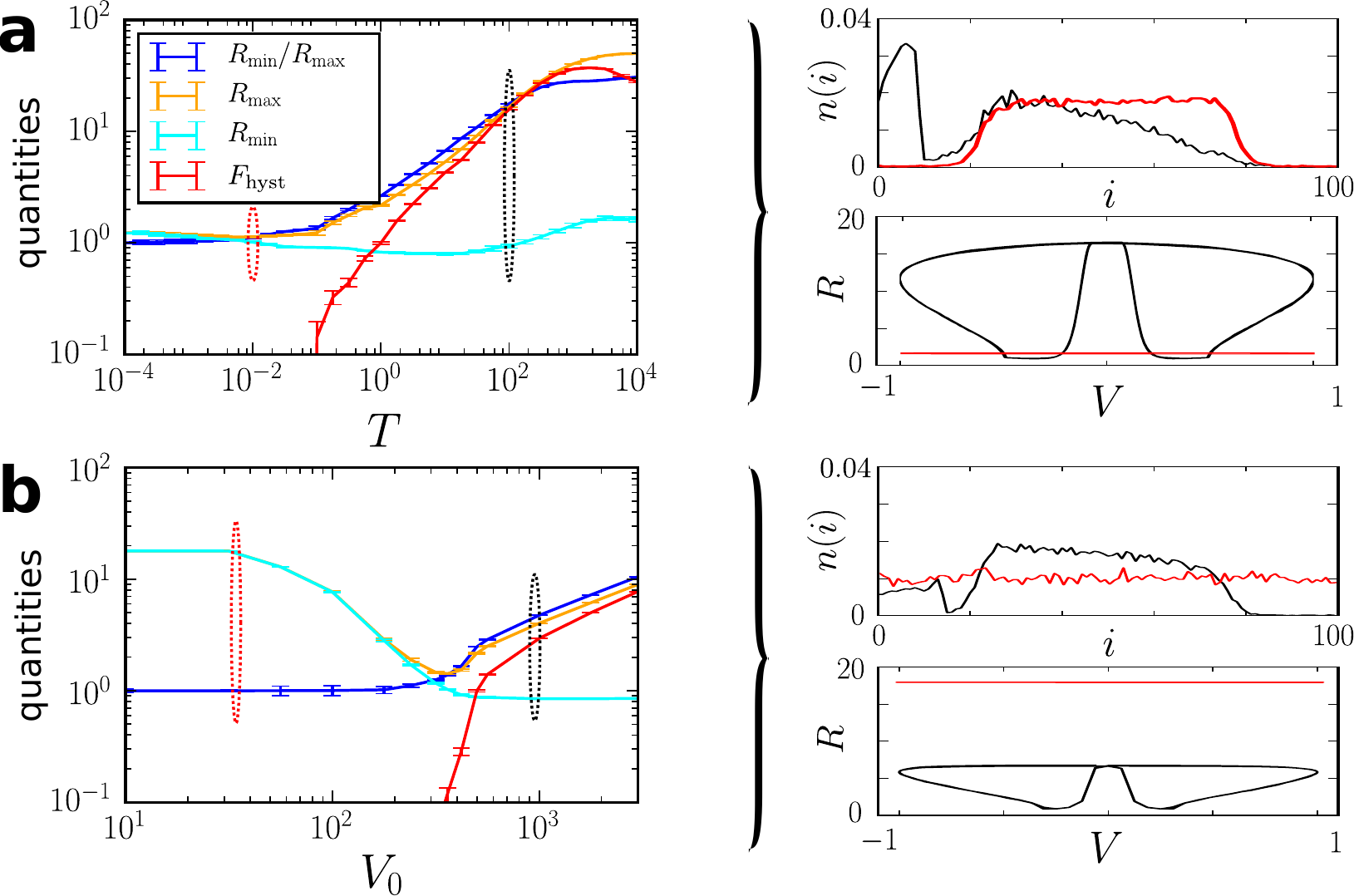}
\caption{Resistance minima and maxima, their respective ratio and hysteresis loops in the $R-V$ diagram, varied over several cycles of sinusoidal voltage driving. The area of the hysteresis loops $F_\text{hyst}$ is normalized to that at $V_0=500, T=1$. (\textbf{a}) Variation of the driving period $T$ at $V_0 = 500$. (\textbf{b}) Variation of voltage amplitude $V_0$ at $T=1$. Shared legend for (\textbf{a}) and (\textbf{b}), all parameters as in \fig{dyn-sine}. The right panel shows sample distributions at $t=0.01$ and resistance over voltage plots for the encircled values of the plot to the left, with curve colors matching the color of the circles.}
\label{variation_T_V0}
\end{center}
\end{figure} 

We further want to investigate the system dependence on the sinusoidal driving, particularly on its period $T$ as well as its amplitude $V_0$. To that end we study two different quantities. Firstly the maximum and minima of the resistance during a switching cycle, $R_\text{min}$ and $R_\text{max}$, and their ratio $R_\text{max}/R_\text{min}$. Secondly, the area of the hysteresis loops in the RV-diagram, $F_\mathrm{hyst}$.

The results are shown in \fig{variation_T_V0}. Therein we see that for very small $T$ the system has no time to react to the sudden changes of the voltage. The minimum and maximum resistances do not differ and the vacancy distribution approaches a state in which, besides fluctuation, every lattice site contributes equally to the resistance, $n_i \propto 1/A_i, R\rightarrow 1$. The corresponding hysteresis loops vanish. For longer periods, $R_\text{max}$ and $R_\text{min}$ start to spread, and the area of the hysteresis loops  increases. This is mainly due to the increment in $R_\text{max}$ when the vacancies accumulate near the interface; however, when the wave reaches the bulk, $R_{\min}$ reaches values considerably lower than in the equilibrium distribution also.

The picture for the amplitudes is somewhat different. Specifically, for very small $V_0$, the system has a high resistance. The diffusive motion that is counteracting the external driving is dominant and pushes the vacancies into a uniform distribution. We have $\m{n_i}_t = 10^{-2}, R \rightarrow \m{n_i}_t \sum A_i \simeq 17.5$. From a certain threshold above $V_0\simeq{75}$, the resistance suddenly drops and the switching dynamics sets in, with corresponding spread in $R_{\text{max}}$ and $R_\text{min}$ and hysteresis loops. Now, the external driving dictates the dynamics. With increasing amplitudes $V_0$ more vacancies can gather in the high resistance regions near the boundaries, which results in a larger maximal resistance $R_\text{max}$. Hence there is a connection between amplitude and period so that the system shows meaningful switching behavior.  

For the relevant time scale, we remark that the transition rates are proportional to the activation energy, $w_{i\rightarrow j} \propto \exp(-E_A)$, cf. \refeq{transprob}. The larger $E_A$ is, the slower all reactions take place. The other element that sets the time scale of the dynamics, is the external voltage $V(t)$. Since we assume a periodically driven system, the length of the period must increase exponentially to offset the delay due to an increased $E_A$. 

\subsection{Effects of Fluctuations}
\label{fluctuations}
\begin{figure}
\begin{center}
\includegraphics[scale=0.7]{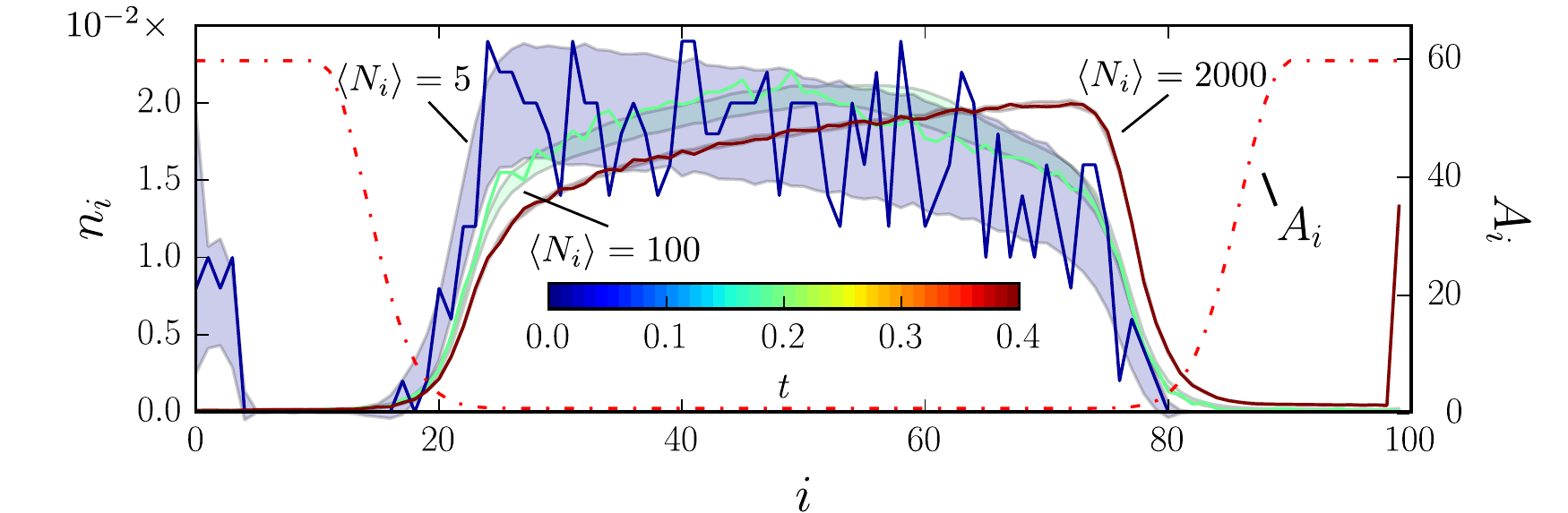}
\caption[Effects of fluctuations on the resistive switching effect]{RS dynamics with enhanced fluctuations due to decreased average vacancy number at constant $\m{N_i}/N_0= 0.01$. Colored faces are the standard deviation around the mean value, inside which one sample distribution for each $\m{N_i}$ is shown. Curves given at different time points of the driving cycle, as indicated by the color scheme.}
\label{fluc_N}
\end{center}
\end{figure}

The influence of stochasticity is investigated by changing the number of vacancies per lattice sites, while keeping the average vacancy density $n_i = \m{N_i}/N_0$ and the other parameters constant. This means, that for a reduced total number of vacancies $\sum_i N_i$ each individual one bears a larger fraction of the total resistance $R$ and hence causes a larger change in the system properties if it hops in between lattice sites with different $A_i$, while the electrical quantities do not differ in principal, since only the ratio $N_i/N_0$ enters (cf. eqs. (\ref{locres}-\ref{volt-drop-disc})). For large vacancy numbers on the other hand, each individual vacancy barely affects the system and its dynamics will approach the deterministic behavior described by the corresponding master's equation \refeq{master-eq}. In \fig{fluc_N} results for various numbers of vacancies per lattice site are depicted at different time points in the switching cycle. Clearly, the fluctuations in the occupation distributions 
increase for smaller $\m{N_i}$.

\begin{figure}
\begin{center}
\includegraphics[scale=0.7]{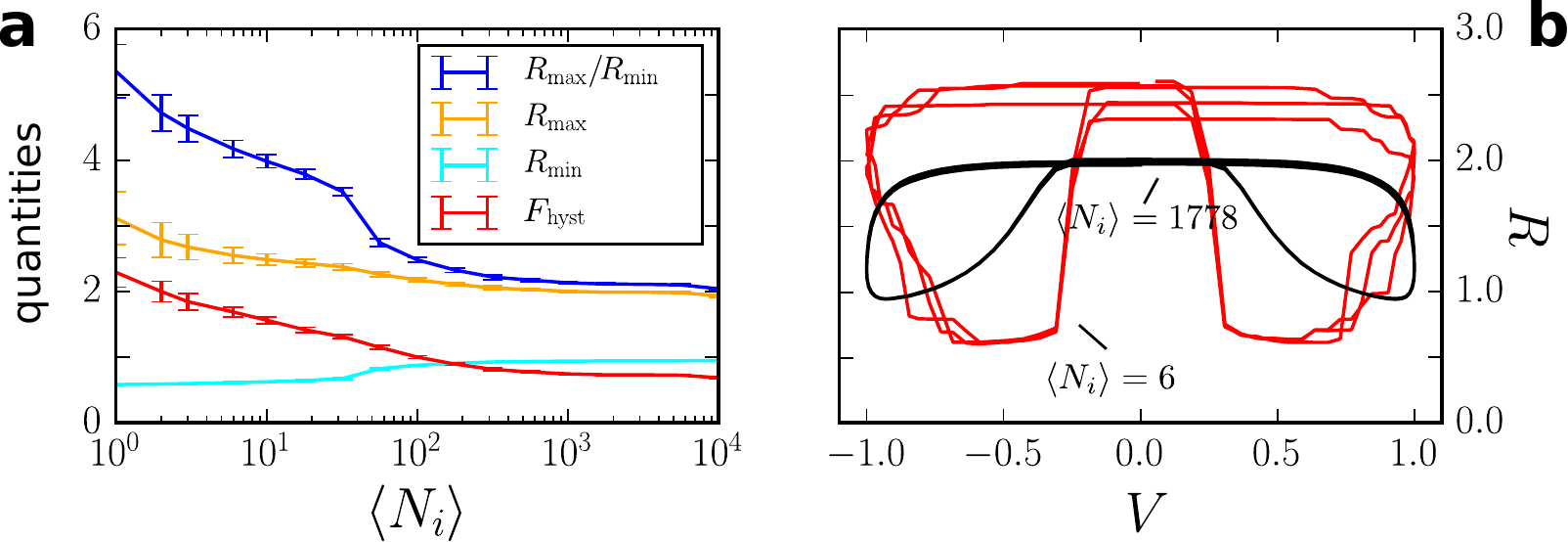}
\caption{Effects due to enhanced stochasticity caused by different mean number of particles per lattice site $\m{N_i}$. (\textbf{a}) Resistance minima and maxima, and hysteresis loops in the R-V diagram as in \fig{variation_T_V0}, albeit as a function of $\m{N_i}$. (\textbf{b}) Sample R-V plots for various $\m{N_i}$. All parameters as in \fig{dyn-sine}. }
\label{fluc_N2}
\end{center}
\end{figure} 

The area of the hysteresis curves of the implicit $V-R$ plot is depicted in \fig{fluc_N2} for varying average particle numbers per lattice site $\m{N_i}$, together with the maxima and minima of the resistance and their respective ratio. We see that for small $\m{N_i}$, the spread in between $R_\mathrm{max}$ and $R_\mathrm{min}$ monotonically increases, which also implies an increase in the size of the corresponding hysteresis loops. Also, we see that the variation between individual cycles are larger for small average particle numbers, which can be seen in form of the larger statistical deviations in this case.

These observations are confirmed in the particle $R-V$ diagram, where we indeed see that both the size of the loops and their fluctuations are bigger for smaller $\m{N_i}$, whereas for many particles fluctuations barely play a role and the results approach the values we gain by direct numerical integration of the corresponding master equation (\ref{master-eq}). Hence we conclude that the resistive switching effect becomes more pronounced with increasing fluctuations. In this sense, the results mirror investigations of the HP-memristor, where it was found that an additional noise increases the memristive response of the system \cite{Stotland2012}.

\subsection{Continuum Limit and Burgers-Like Equation for Wave Transit}
\label{sectionburgers}
In this part, we want to introduce a continuum limit of the master equation in a mean field approximation (MFA). Here, only the outline of the calculations is given. A detailed derivation is shown in the appendix \ref{app:burgers}. For the time derivative of the mean occupation density at lattice site $i$, cf. \refeq{particledensity}, we have
\begin{align}
\begin{aligned}
\frac{\mathrm{d}}{\mathrm{d}t} \m{n_i(t)} = 
 + \m{n_{i-1} w_{i-1\rightarrow i}}  - \m{n_i w_{i \rightarrow i+1}} 
 + \m{n_{i+1} w_{i+1 \rightarrow i}}  - \m{n_i w_{i \rightarrow i-1}} .
\end{aligned}
\label{nk-prime}
\end{align}
This expression follows from the master equation (\ref{master-eq}) and is derived in the appendix. Now we decouple the two point and higher order correlations, resulting in 
\begin{align}
\begin{aligned}
\m{n_i w_{i\rightarrow j}} =  \m{n_i} \lef 1-\m{n_j} \rig e^{-E_A}\m{e^{\Delta V_i}} 
\label{mfa-hop-discrete-1}
\end{aligned}
\end{align}
Next, we introduce the lattice spacing $\epsilon$ between neighboring sites and define $x :=i \epsilon$. Formally, the continuum limit is taken by letting the lattice spacing become infinitesimal while the number of lattice sites $N$ runs to infinity, in such a way that the product of both remains the constant lattice length $\lim_{\epsilon \rightarrow 0, N\rightarrow \infty} \epsilon N = L $.  The hopping rates that formerly depended upon the density of neighboring lattice sites now depend on those at an infinitesimal distance $\epsilon$, which will obviously introduce a derivative. The whole approach is quite similar to the derivations of Burgers equations as found for molecular motors in the ASEP model \cite{Ferrari1992}. From here, we will use the rescaled length $\tilde x = x/L$ and drop the tilde. 

Let the averaged density profile be denoted by $\rho(x)$, from the connection 
 \begin{equation}
\m{n_i}(t) = \int_{i\epsilon}^{(i+1)\epsilon}\dd x\rho(x,t) 
\label{rho_connection}
 \end{equation}
 we obtain $\epsilon \rho(x,t)=\m{n_i}(t)$. Recall that our previous choice of a lattice with $100$ sites corresponds to $\epsilon = 0.01$.   
The continuous voltage $V(x)$ is introduced analogously to \refeq{rho_connection}. Entering this into the MFA hopping rate \refeq{mfa-hop-discrete-1}  and developing the resulting expression to the second order  yields the equation
\begin{align}
\begin{aligned}
\frac{\partial}{\partial t} \rho(x,t)  =  e^{-E_A}\cosh\lef\partial_x V\epsilon\rig 	\epsilon^2\partial^2_x \rho		
	-2e^{-E_A}\sinh\lef\partial_x V\epsilon\rig 	\epsilon \partial_x \rho.
 \label{protoburger}
 \end{aligned}
 \end{align}
As the electrical current through the system is not position dependent, we express local voltage in the following way (cf. \refeq{volt-drop-disc})
\begin{equation}
\partial_x V(x,t) = \frac{V(t)}{\int \dd x' A(x')\rho(x',t)} A(x)\rho(x,t) =: I(t) A(x)\rho(x,t),
\end{equation}
thereby splitting it in a nonlocal and local component. Bearing in mind the continuity equation $\partial_\tau \rho + \mathrm{div} j = 0$, \refeq{protoburger} is recast into
\begin{align}
\begin{aligned}
\frac{\partial}{\partial t} \rho(x,t) = \partial_x\big[e^{-E_A}\cosh\lef I A \epsilon \rho\rig 	\epsilon^2\partial_x \rho\big] 
- \partial_x \big[2e^{-E_A}\sinh\lef I A \epsilon \rho \rig  \epsilon\rho\big] ,
 \end{aligned}
 \end{align}
Here, the term in braces is the oxygen vacancy flux. Since $\rho$ can be considered a probability density with the normalization $\int_0^1 \dd x \rho(x)=1$, this constitutes nonlinear continuity equation for the probability. By rescaling the time, the prefactor $\epsilon^2 \exp(-E_A)$ can be eliminated. By further, expanding the hyperbolic function to the first order, we are left with
\begin{equation}
\frac{\partial}{\partial t} \rho(x,t) = \partial_x^2 \rho(x,t) - 2 I(t) \partial_x \lef A(x)\rho^2(x,t) \rig.
\label{burgers-2}
\end{equation}
This equation can be formally compared to the viscous  Burgers equation. With the viscosity denoted by $\nu$, it reads $\partial_t u =  \nu\partial_x^2 u - 1/2 \partial_x u^2 $. So \refeq{burgers-2} is a Burgers-like equation with a current and resistance profile dependent prefactor. Following the wave-like motion of the vacancies, this connection has also been noted in \cite{Tang2014}.

\begin{figure}[h]
\begin{center}
\includegraphics[scale=.7]{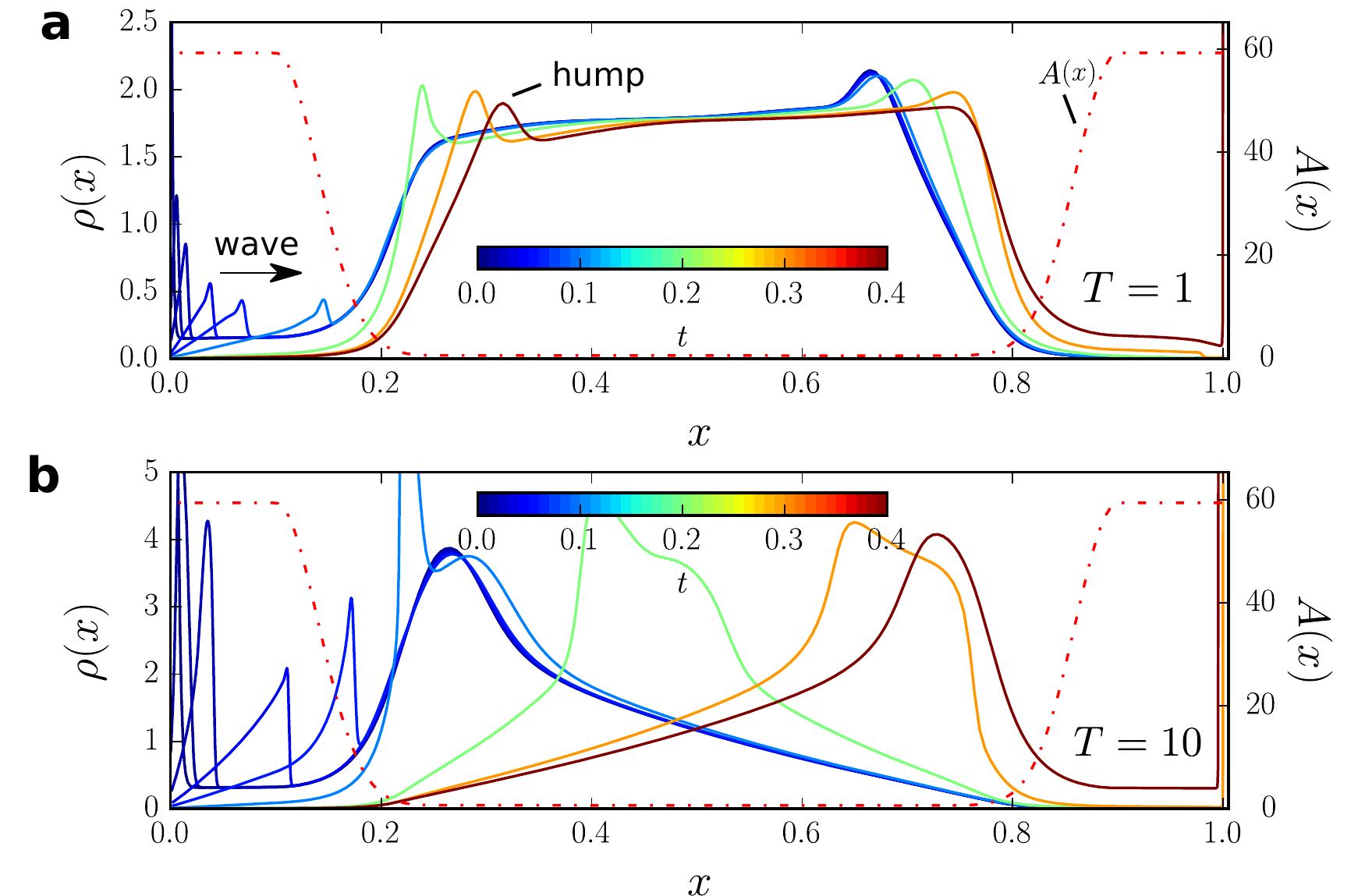}
\caption{Vacancy distribution of resistive switching modeled by numerical integration of the corresponding Burgers equation (\ref{burgers-2}). Various time points of the switching cycle indicated by the color-coding. Parameters as in Fig.\ \ref{dyn-sine}, for driving periods (\textbf{a}) $T=1$ and (\textbf{b}) $T=10$. }
\label{rho-burgers}
\end{center}
\end{figure}

To validate the derived result, we aim at numerically integrating \refeq{burgers-2} and compare the thereby obtained results with the original numerical results. The integration is done with Neumann boundary conditions, there is no density flux out of the system. The traveling wave solution is shown in \fig{rho-burgers}. Similar to the results seen in Fig. \ref{dyn-sine}.\textbf{b}, at the beginning of a period there is a surplus of vacancies stored at the left hand side of the RS-device, near $x=0$,  which increases significantly with the driving period $T$. Upon being agitated by a voltage, these vacancies set in motion in form of a wave with decreasing peak height, until it reaches the bulk. For $T=1$ the distribution in the bulk is almost flat, the incoming wave forms a hump that moves further to the right with time. Meanwhile, before the second bulk interface around $x=0.7$ there is initially a hump that flattens and finally disappears as the distribution shifts further to the right. For $T=10$ on the 
other hand,  the initial distribution in the bulk looks like an inclined plane. The incoming wave rushes through the bulk here, reversing the orientation of the inclined plane on its way through it.  

\begin{figure}[h]
\begin{center}
\includegraphics[scale=.7]{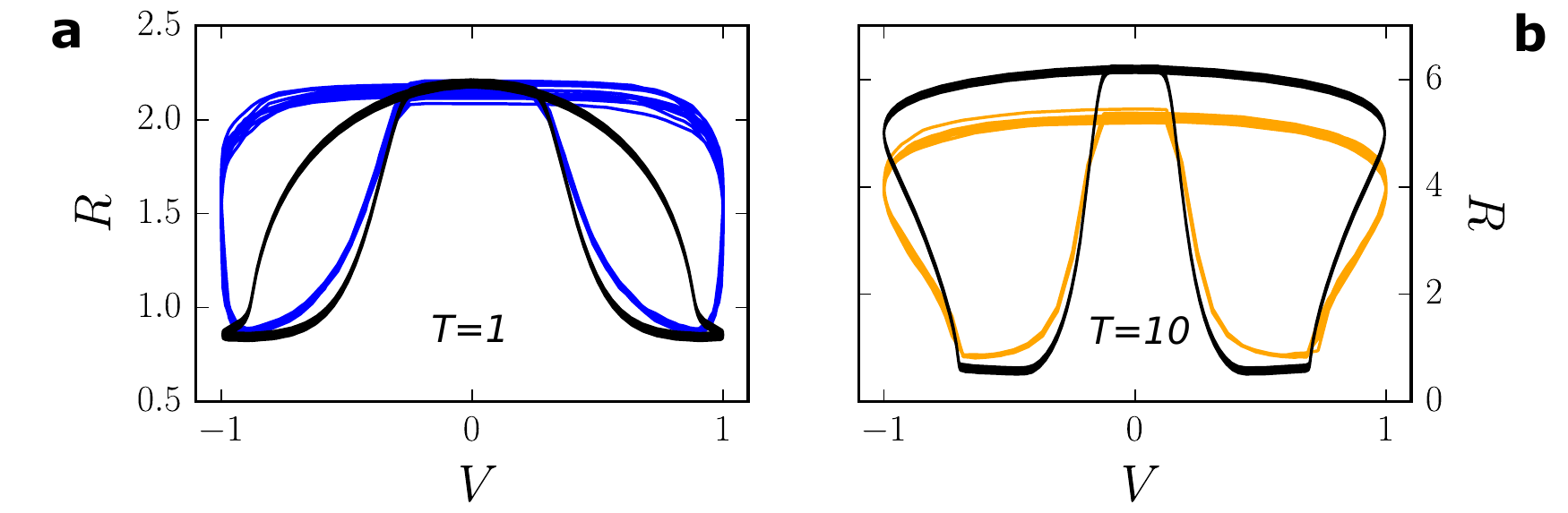}
\caption{Resistance over voltage of particle simulation (colored, cf. \refeq{master-eq}) vs numerical integration of Burgers equation (black, cf. \refeq{burgers-2}) for several cycles of the driving periods (\textbf{a}) $T=1$ and (\textbf{b}) $T=10$.}
\label{VR-burgers}
\end{center}
\end{figure}

As we can see, the traveling wave reproduces the previous results with good principal agreement, as shown exemplary in the $R-V$ for two different driving periods in Fig. \ref{VR-burgers} superimposed on the corresponding diagrams gained by integration of the particle picture. For all driving periods $T$, we see some alteration of the two-legged structure. 

\section{Logical States} 
\label{sectionlogic}
\begin{figure}
\begin{center}
\includegraphics[scale=0.9]{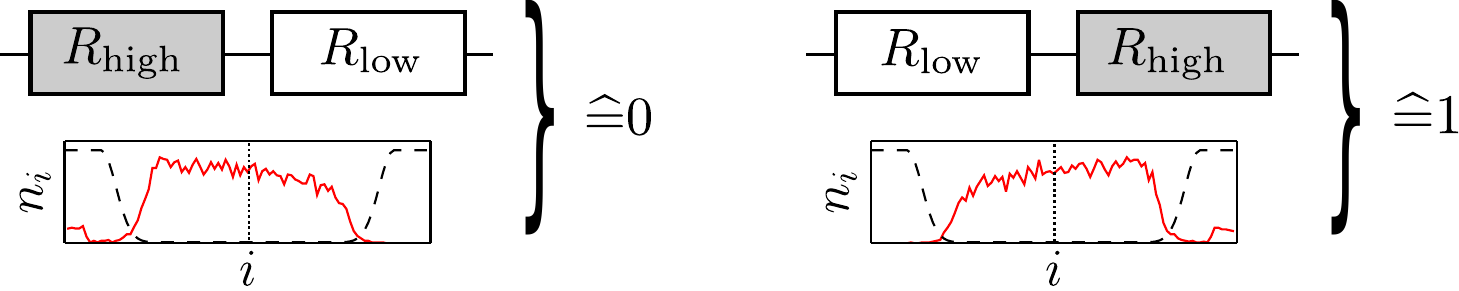}
\caption{Logical states of the RS device. 
The top panels illustrates how the lattice can be thought of as two serial resistors, one in a high and one a low resistive state. Their order then determines the logical state. The bottom panels feature a sample vacancy distribution (red color) in the lattice according to the respective states, together with the dashed resistivity profile $A_i$. A further dotted line cuts the lattice in two halves, each of which corresponds to one of the serial resistors above. An accumulation of vacancies near the left side results in a high resistance of that part, and vice versa with a mirror inverted accumulation near the right side.}
\label{setup}
\label{logicalstates}
\end{center}
\end{figure}

As we have seen in the previous sections, the resistance of the system depends on how many vacancies are stored within the bulk versus in one of the two boundary regions. Principally, this configuration is similar to two serial switches with counter oriented polarities, and one active switching zone each. To that end, we think of the lattice as cut in half and each site constituting one resistor that can be either in a high resistive or low resistive state, depending on the distribution of vacancies. This also determines the logical state of the device, as indicated in  \fig{logicalstates}. A gathering of vacancies near one of the electrodes, i.e. in an area with $A_i \gg 0$, leads to a large resistance, whereas in the bulk vacancies only marginally contribute to the resistance, since here $A_i\simeq 0$. Hence, a gathering of vacancies near the first electrode-TMO interface corresponds to a setting $R^{(1)}_\text{high} \oplus R^{(2)}_\text{low} \widehat{=} 0$, the logical zero, and a gathering near the 
second interface corresponds to $R^{(1)}_\text{low} \oplus R^{(2)}_\text{high} \widehat{=} 1$, the logical one. Other configurations, namely an on-on state,  $R^{(1)}_\text{low} \oplus R^{(2)}_\text{low}$, can occur during a read or in the transient of a switching operation, but has no logical equivalent, whereas a vacancy gathering near both electrode interfaces, i.e.  $R^{(1)}_\text{high} \oplus R^{(2)}_\text{high}$ does not occur during the operation of the device \cite{Linn2010}.

\subsection{Reading and Writing Operations}
\label{sectionreadingop}
Now that we have defined the logical states, we can express the actions of the driving voltage in these terms and develop a framework of how to write and read information in our system.

\begin{figure}
\begin{center}
\includegraphics[scale=.8]{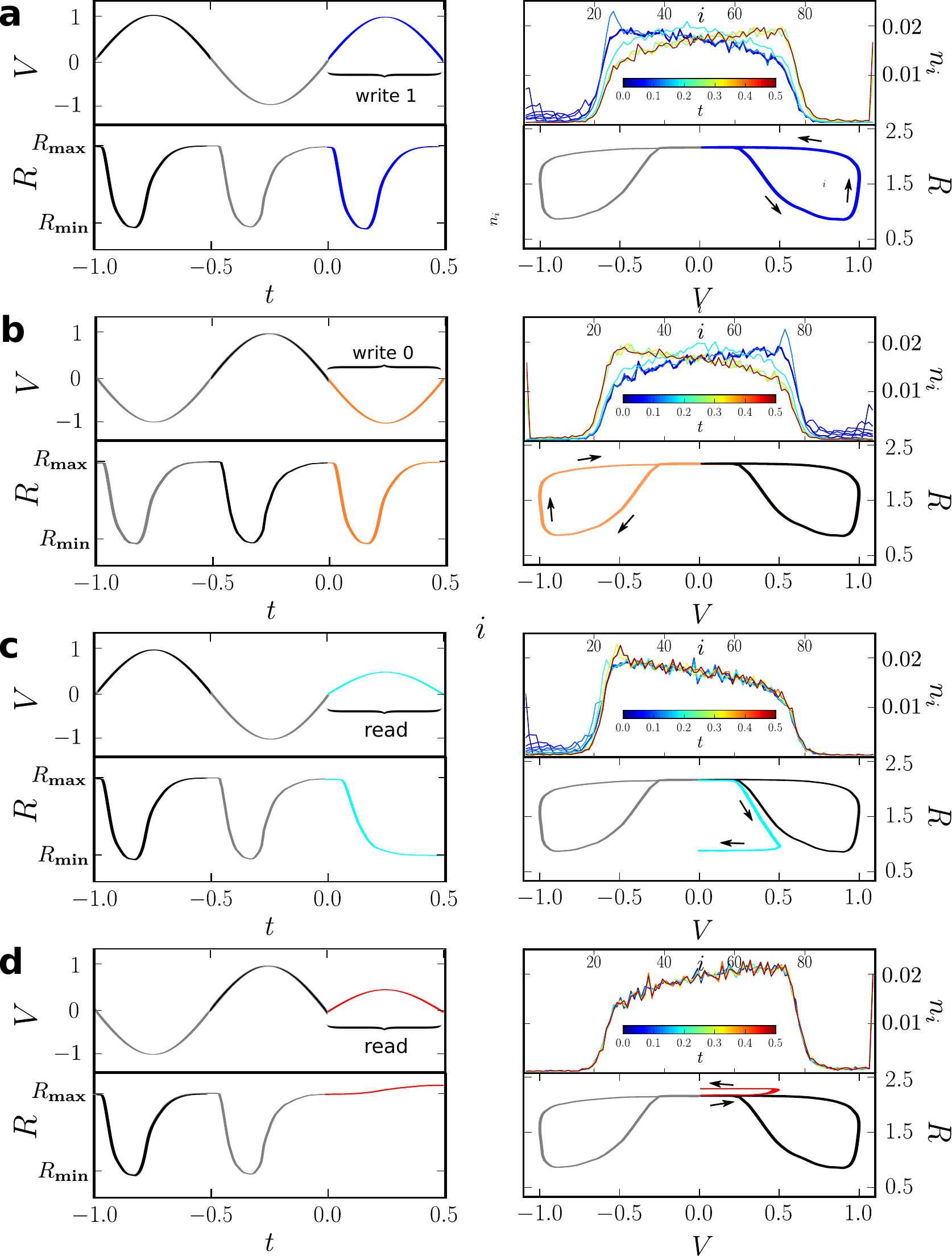}
\caption{Writing and reading operations in the stochastic RS device. The left panels depict the resistance and voltage over time diagrams, and the right panels depict vacancy distributions at several time points of the specific operation as indicated by the color coding; and the implicit $R$ over $V$  plot.  (\textbf{a}) System prepared in state $0$ by several writing cycles $0 \rightarrow 1 \rightarrow 0$, then write $1$ operation. (\textbf{b}) System prepared in state $1$ by several writing cycles $1 \rightarrow 0 \rightarrow 1$, then write $0$ operation. (\textbf{c}) System prepared in state $0$, then reading operation. (\textbf{d}) System prepared in state $1$, then reading operation. Writing operations with $V_0 =500$, reading operation with $V_0 =250$, all other parameters as in \fig{dyn-sine}. }
\label{write-read}
\end{center}
\end{figure}

In \fig{dyn-sine} we have seen how the positive half-cycle of the external voltage driving acts upon an initial vacancy accumulation near the left interface. Namely, by pushing it to the right interface. During its course, the resistance drops when the vacancy wave hits the bulk, but returns to its initial value once the voltage pulse ends and enough vacancies have accumulated near the right interface. In terms of the logical states, this corresponds an initial $0$-state, onto which the positive voltage pulse acts. Afterwards, the system is in a $1$-state. Hence the process is the write $1$ operation.  It is again displayed in \figa{write-read}{a}.

In \figa{write-read}{b} the antagonistic process is shown, where the negative voltage half-cycle acts on a $1$-state, resulting in a $0$-state.  Mirror reversed, the vacancies move from the right interface to the left one. Hence the negative voltage pulse performs the opposing write $0$ operation. During its course, the resistance has the same dynamics. We note that the initial states here and for the read operation are prepared by several switching cycles $0\rightarrow1\rightarrow0\rightarrow\dots$. 

Let us turn to the reading operation. The state of the system is determined by sending a reading pulse through it and monitoring the reaction. The reading pulse itself is is the positive voltage half-cycle, albeit with half the amplitude of the writing signal $V_\text{read} = V_0/2$.

If the reading pulse acts on an initial $0$-state, the resistance drops to its minimum level, $R_\text{min}$. For the oxygen vacancy distributions this we see that the initial vacancy accumulation near the left interface dissolves and moves to the bulk. But unlike for the write case, there is no subsequent buildup of vacancies near the right interface. Hence when the reading pulse has passed, the system remains in the low resistive state $R^{(1)}_\text{low} \oplus R^{(2)}_\text{low}$. The process is shown in \figa{write-read}{c}.

On the other hand, if the system is prepared in the $1$-state, the reading pulse leads to a slight increase in the resistance above the value $R_\text{max}$ that is obtained for the switching cycle $1\rightarrow 0 \rightarrow 1$. For the distribution, there are virtually no vacancies near the left interface that can be transported to the bulk, while the accumulation near the right electrode already exists and only slightly increases in size. Hence in this case we have only a small effect on the resistance; cf. \figa{write-read}{d}.

This means that two logical states are distinguished by their differing reactions to the reading pulse. Only when it acts upon the $0$-state there is a distinct drop in the resistance. Obviously, the reading pulse has changed the initial distribution, hence a reversely oriented reset pulse need to restore the original configuration in either case.

\subsection{Clarity and Stability of the Reading Operation}
\label{sectionclarity}
As noted in section \ref{fluctuations}, the resistance drop increases with lower particle numbers and so do the statistical fluctuations associated with it.  To be able to correctly determine the state of the device, we must assure that the respective resistances after a read pulse and the corresponding reset pulse do not overlap.  This property is demanded for a single read-reset cycle as well as for several. 

\begin{figure}
\begin{center}
\includegraphics[scale=.8]{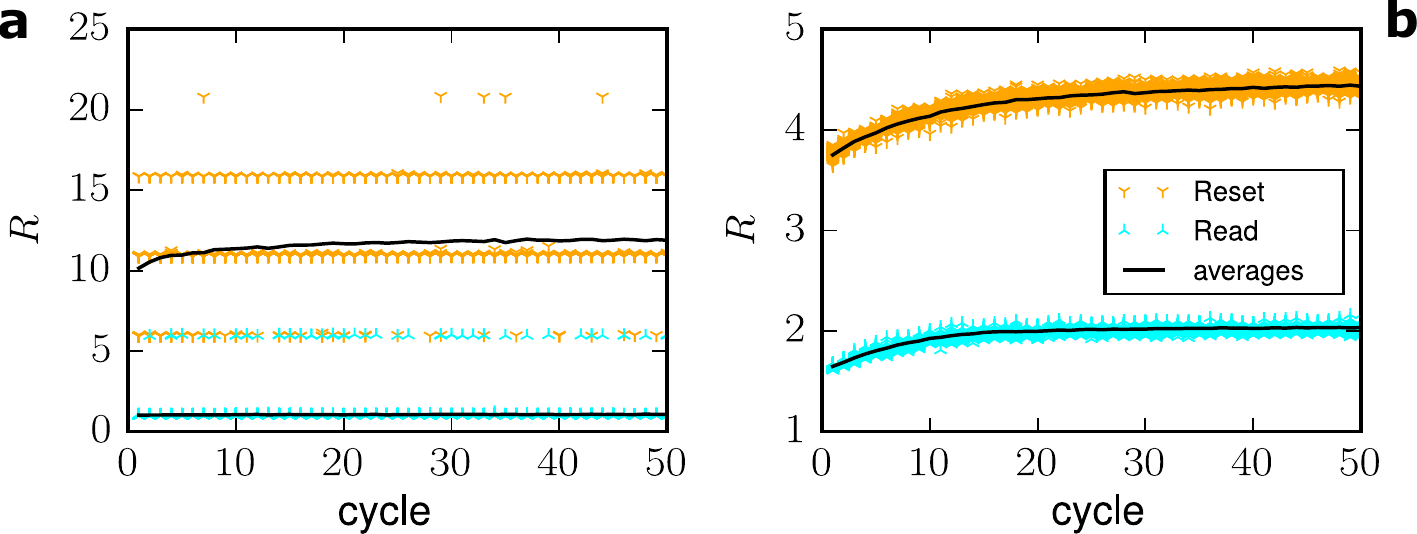}
\caption{Resistances after several repeats (=cycles) of read and reset operations, for different average particle numbers per lattice site, (\textbf{a}) $\m{N_i}=0.2$  and (\textbf{b}) $\m{N_i}=178$. Other parameters as in \fig{dyn-sine}.}
\label{read-reset-cycles}
\end{center}
\end{figure} 
\begin{figure}
\begin{center}
\includegraphics[scale=.8]{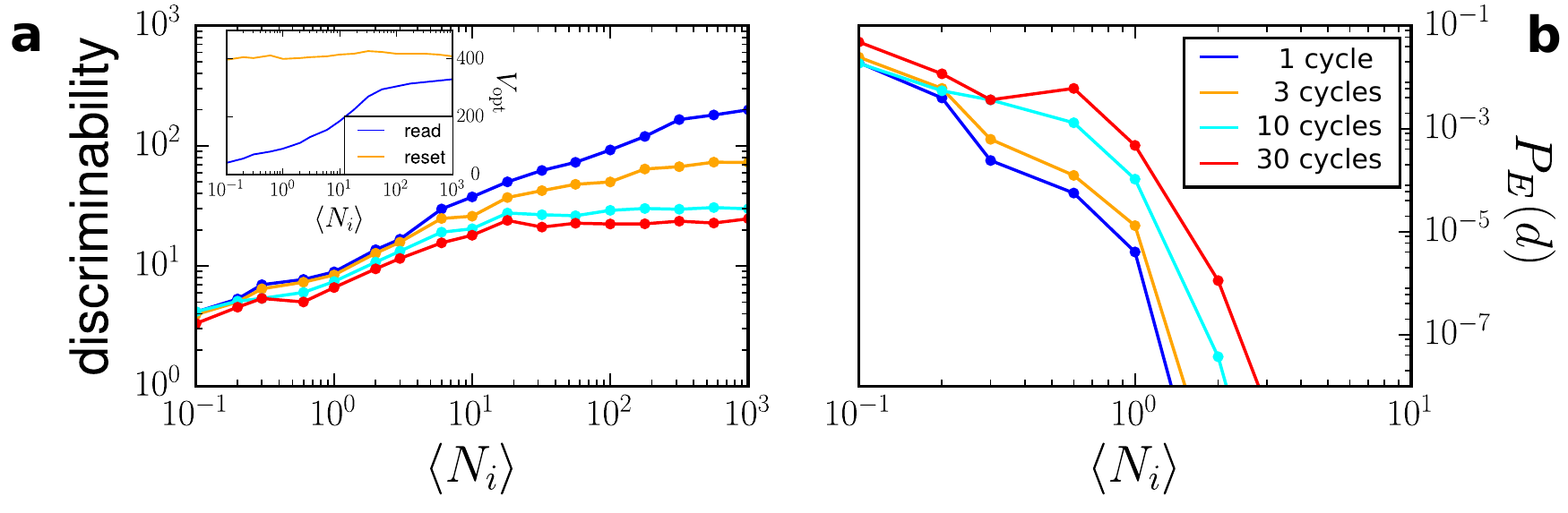}
\caption{Discriminability and associated probability of a reading error as a function of the average particle number for various numbers of read-reset-cycles. The inset denotes the optimal reading and reset voltages, i.e., those for which the reading resistance drops to a minimum and the reset resistance is close to the value reached in the writing cycle, respectively. Other parameters as in \fig{dyn-sine}.}
\label{discriminability}
\end{center}
\end{figure} 

In \fig{read-reset-cycles} various such cycles are shown for two different average particle densities $\m{N_i}$. As we can see, for $\m{N_i}=0.2$, the possible outcomes for the resistances fall almost perfectly onto discrete levels, one of which is overlapping. Here, we cannot determine in which state the system actually is. For ${\m{N_i}=178}$ on the other hand, the ratio of $R_\mathrm{reset}/R_\mathrm{read}$ is much smaller, but so are the fluctuations accompanying them. Hence both states can be distinguished easily.

In order to quantify the effect, we employ the measure of the discriminability. Originally defined as the difference between the signals with and without noise \cite{Engel2009}, we adapt it to our current situation by setting
\begin{equation}
    d = \frac{| \m{R_\mathrm{reset}} - \m{R_\mathrm{read }}|}{\sigma}
\end{equation}
with the standard variations of the respective resistances denoted by $\sigma$. As such, it is closely related to the signal to noise ratio, which simply is its square: $SNR = d^2$. We approximate that the results fall onto a normal distribution, with peaks at $\m{R_\mathrm{reset}}$  and $\m{R_\mathrm{read }}$ and the standard deviation $\sigma$. This is justified since we are interested only in an estimate of the error. 

In \fig{discriminability} this measure is depicted as a function of the average particle number $\m{N_i}$. With increasing $\m{N_i}$, the discriminability of the output improves, since fluctuations largely vanish. Moreover, it drops with the number of read-reset cycles, a spread which is more pronounced for large $\m{N_i}$. In view of \fig{read-reset-cycles}, this can be attributed to the transient of the average resistance values. The optimal $V_\text{read/reset}$ (shown in the inset of \fig{discriminability}) was determined after a writing operation,  $1 \rightleftharpoons 0 \rightarrow \text{read}$, and apparently differs slightly from the value that is obtained after a read/reset cycle, $\text{read}\rightleftharpoons\text{reset}\rightarrow\text{read}$, thereby increasing the interval of the possible outcomes. 

To interpret the discriminability measure, we check on how many values will be falsely attributed. The measurements of the read and reset resistances fall onto two normal distributions whose peaks are $ \sigma\cdot d$ apart. All measured values  $x \leq R_\mathrm{read}+\sigma d /2 =: x_\mathrm{mid}$ are then assigned to the distribution of $R_\mathrm{read}$, all larger values to $R_\mathrm{reset}$.
Hence the probability of an erroneous attribution is given by those reset values smaller than $\m{R_\mathrm{read}}+\sigma d/2$ plus the reading values larger than it. Expressed as a formula
\begin{align}
\begin{aligned}
P_E =
&\frac{1}{2}\frac{1}{\sqrt{2\pi\sigma^2}} \int_{x_\mathrm{mid}}^\infty dx e^{-(x-\m{R_\mathrm{read}})^2/2\sigma^2} 
 + \frac{1}{2}\frac{1}{\sqrt{2\pi\sigma^2}} \int_{-\infty}^{x_\mathrm{mid}} dx e^{-(x-\m{R_\mathrm{reset}})^2/2\sigma^2} 
 = \frac{1}{2}\mathrm{erf}\left(\frac{d}{2\sqrt{2}}\right),
\end{aligned}
\end{align}
where erf$(x)$ denotes the error function. The corresponding results are plotted in \fig{discriminability}. We conclude that for an average vacancy occupation per lattice site of $\m{N_i} \geq 10$ we are secure independently of the number of cycles. Hence, the unambiguity of the reading establishes a lower boundary for the optimal particle number, while the clarity of the resistive switching effect is emphasized with lower particle numbers. This favors a intermediate range of the optimal average particle number of about $N_\mathrm{optimal} = \sum_i N_i = 100\cdot \m{N_i} =1000$ particles.

We note that the number of vacancies scales with the size of the device, hence the risk of erroneous read-outs limits the possible degree of miniaturization. In a rough estimate, we find that this translates into a length of about $18\mathrm{nm}$, assuming a cubical element of lanthanum-strontium manganite  $\mathrm{La}_{1-x}\mathrm{Sr}_x\mathrm{MnO}_{3-x/2}$  with a vacancy density of $x=0.3$ \cite{Trukhanov2004}.

In actual systems, there is intrinsic asymmetry between the upper and lower electrodes, which can be reflected by different heights of the left and right flank of the resistivity profile $A_i$ in our system. This will result in two different  `leg' lengths in the resistance over voltage diagrams, but does not change the stochastic fluctuations, i.e. the results for the discriminability apply to this scenario as well.

\section{Conclusion}
Resistive switching is a topical and high current interest field in condensed matter physics and a prototype of a nonlinear stochastic switching event. Its main application, the non volatile resistive random access memory, is a promising candidate for future memory technologies. The scaling to small elements however, leads to an increasing role of fluctuations. It is therefore required, to have an understanding of the switching mechanism and the reliability under the influence of these fluctuations. 
Here we develop a particle based mesoscopic model based on the distribution of oxygen vacancies. Fluctuations caused by the stochasticity of their motion both enhance the resistive contrast and reduce the reliability of the resistive switch. These counteracting tendencies enable us to predict a limit to miniaturization and an optimal system size.

We have introduced a setup to describe a complementary resistive memory switch based on a discrete particle hopping model. Hereby, the spatial distribution of oxygen vacancies plays the vital part in determining the state of the system, which switches between a high and low resistive state. The vacancies' dynamics is given by a master equation with transition rates depending on the vacancy distribution and an external voltage driving. 

The application of voltage pulses led to an accumulation of vacancies near the electrode-TMO interface. An antagonistic pulse then sets these accumulations in motion and they collectively wander through the system, thereby affecting the resistance. By formulating the problem in a particle picture, we gained a tool to vary the accompanying fluctuations: less vacancies, more fluctuations. We looked at the characteristics that define an RS element, such as the spread between the high and low resistance, and discovered that they become more pronounced with increasing fluctuations. 

The nature of this collective vacancy motion could be elucidated by deriving a nonlinear continuity equation in a MFA from the master equation in continuum space. It has the structure of a Burgers equation with an additional nonlocal factor, that also incorporates the influence of the driving. We succeeded at numerically integrating this equation to find good agreement with the results gained by particle simulations. Hence we interpreted the motion of the vacancies as a nonlinear traveling wave. 

Further, we defined binary logical states in terms of the underlying particle distributions. By linking the actions of voltage pulses to switches between these states, we have established the reading and writing operations necessary to run a memory element. Interestingly, investigations in the stability and discriminability of these operations let us gain a lower limit for the possible number of vacancies in the system, a quantity that is connected with the possible level of miniaturization. Together with the finding of enhanced RS-characteristics for fewer vacancies, this results in an optimal performance for about 1000 oxygen vacancies, which corresponds to a device length of about $18$nm.

\section*{Acknowledgements}
This paper was developed within the scope of the International Research Training Group 1740 funded by the Deutsche Forschungsgemeinschaft. ALH and AVS acknowledge the Visiting Fellowships by the Isaac Newton Institute for Mathematical Sciences, Cambridge, UK for setting up a collaborative link within the CFM Program, ``Mathematical Modeling and Analysis of Complex Fluids and Active Media in Evolving Domains'', 2013.

\section*{References}

\bibliography{../bibliography/library_manual}

\begin{appendix}
\section{Derivation of the Burgers' Equation}
\label{app:burgers}
In this appendix, we derive a wave equation from the master equation that governs our system. We proceed in two steps: at first, we manipulate the discrete Master equation to gain the time derivation of the average occupation of a lattice site. With this result, we then introduce the continuum limit and expand it to the second order. In the master equation
\begin{align}
&\frac{\mathrm{d}}{\mathrm{d}t}P(\mathcal{N},t) =
\sum_{\mathcal{\{N'\}}\neq\mathcal{N}}\Big( -W \lef \mathcal{N}\rightarrow \mathcal{N'}\rig P(\mathcal{N},t)+ W \lef \mathcal{N'}\rightarrow \mathcal{N}\rig P(\mathcal{N'},t)\Big),
\end{align}
a configuration of lattice occupations is denoted by $\mathcal{N} = (N_1, N_2, \cdots, N_L)$ and $\{\mathcal{N}\}$ is the set of all possible configurations. With it,  we can express the time derivative of an individual lattice site occupation $\dot{\m{N_k}}$ as
\begin{align}
 \frac{\mathrm{d}}{\mathrm{d}t}\m{N_k} & = \frac{\mathrm{d}}{\mathrm{d}t} \sum_{\mathcal{\{N\}}}N_k P(\mathcal{N},t) \\
& =\sum_{\mathcal{\{N\}}} N_k \sum_{\mathcal{\{N'\}}\neq\mathcal{N}}\Big( -W \lef \mathcal{N}\rightarrow \mathcal{N'}\rig P(\mathcal{N},t)+ W \lef \mathcal{N'}\rightarrow \mathcal{N}\rig P(\mathcal{N'},t)\Big).
\label{nk-1}
\end{align}
Since only hopping between neighboring sites is allowed, the number of possible transistions is vastly reduced. By $\mathcal{N'} = (\cdots, N_i', N'_{i+1}, \cdots)$ we denote a configuration, that differs from $\mathcal{N}$ only at the $i$-th and $i+1$-th position, where it takes the values $N_i'$ and $N'_{i+1}$ respectively. The remaining transitions of the second sum of \refeq{nk-1} can now be written explicitly, yielding
\begin{align}
\begin{aligned}
\frac{\mathrm{d}}{\mathrm{d}t}\m{N_k} =   \sum_{\mathcal{\{N\}}}N_k\Bigg[ 
& -\sum_{i=2}^L W \lef \mathcal{N} \rightarrow \dots, N_{i-1}+1, N_i-1, \dots \rig P(\mathcal{N},t) \\
&  -\sum_{i=1}^{L-1} W \lef \mathcal{N} \rightarrow \dots, N_{i}-1, N_{i+1}+1, \dots \rig P(\mathcal{N},t)\\
 +\sum_{i=2}^{L}    W &\lef \dots, N_{i-1}+1, N_i-1, \dots \rightarrow \mathcal{N} \rig P(\dots, N_{i-1}+1, N_i-1, \dots,t) \\
 +\sum_{i=1}^{L-1} W & \lef \dots, N_{i}-1, N_{i+1}+1, \dots \rightarrow \mathcal{N} \rig P(\dots, N_{i}-1, N_{i+1}+1, \dots,t)\Bigg]\\
\label{nk-ref} 
\end{aligned}
\end{align}
We seek to collect the terms in the brackets by performing a substitution $N_i \pm 1 =  N'_i$, such that the probability function depends on a configuration $\mathcal{N'}$ that is not altered by subtraction or addition of a particle. This substitution can either affect $N_k$ or not. First, the terms of \refeq{nk-ref} without particle hop at the $k$-th position:
\begin{align}
\begin{aligned}
-\sum_{\mathcal{\{N\}}}N_k	 \Bigg[ &  \sum_{i\neq k, k+1} W \lef \mathcal{N} \rightarrow \dots, N_{i-1}+1, N_i-1, \dots \rig P(\mathcal{N},t) \\
						 +&\sum_{i\neq k-1,k} W \lef \mathcal{N} \rightarrow \dots, N_{i}-1, N_{i+1}+1, \dots \rig P(\mathcal{N},t)\Bigg] \\
+\sum_{\mathcal{\{N'\}}}N_k \Bigg[ &\sum_{i\neq k-1,k} W \lef   \mathcal{N'} \rightarrow \dots, N'_{i}+1, N'_{i+1}-1, \dots \rig P(\mathcal{N'},t) \\
						+& \sum_{i\neq k, k+1} W \lef \mathcal{N'} \rightarrow \dots, N'_{i-1}-1,  N'_i+1, \dots \rig P(\mathcal{N'},t) \Bigg]  = 0
\end{aligned}
\end{align}
Since the summation is done over all possible configurations in both sums, we can rename $\mathcal{N'}$ to $\mathcal{N}$ and see that all  summands vanish. In the remaining terms the hopping involve the particles at the $k$-th position, and hence substitutions affect the prefactor $N_k$ in \refeq{nk-ref},
\begin{align}
\begin{aligned}
\frac{\mathrm{d}}{\mathrm{d}t}\m{N_k} = \sum_{\mathcal{\{N\}}}N_k\bigg[   &
- W \lef \mathcal{N} \rightarrow \dots, N_{k-1}+1, N_k-1, \dots \rig P(\mathcal{N},t) \\
&
- W \lef \mathcal{N} \rightarrow \dots, N_{k}+1, N_{k+1}-1, \dots \rig P(\mathcal{N},t) \\ 
& 
- W \lef \mathcal{N} \rightarrow \dots, N_{k-1}-1, N_{k}+1, \dots \rig P(\mathcal{N},t) \\
&
-  W \lef \mathcal{N} \rightarrow \dots, N_{k}-1, N_{k+1}+1, \dots \rig P(\mathcal{N},t)\bigg] \\
+\sum_{\mathcal{\{N'\}}}\bigg[  
& 
(N'_k-1)W \lef \mathcal{N'} \rightarrow \dots, N'_{k}-1,  N'_{k+1}+1, \dots \rig P(\mathcal{N'},t) \\
& +
(N'_k+1)W \lef \mathcal{N'} \rightarrow \dots, N'_{k-1}-1,  N'_k+1, \dots \rig P(\mathcal{N'},t) \\
& +
(N'_k+1)W \lef   \mathcal{N'} \rightarrow \dots, N'_{k}+1, N'_{k+1}-1, \dots \rig P(\mathcal{N'},t) \\
& +
(N'_k-1)W \lef   \mathcal{N'} \rightarrow \dots, N'_{k-1}+1, N'_{k}-1, \dots \rig P(\mathcal{N'},t) 
 \bigg] .
\end{aligned}
\end{align}
Relabeling the marked quantities $\mathcal{N'}$ to unmarked ones shows that all expressions with prefactor $N_k$ cancel each other out. This leaves us only with
\begin{align}
\begin{aligned}
 \frac{\mathrm{d}}{\mathrm{d}t}\m{N_k} =  \sum_{\mathcal{\{N\}}} & \bigg[ 
 	-W \lef   \mathcal{N} \rightarrow \dots, N_{k-1}+1, N_{k}-1, \dots \rig P(\mathcal{N},t) \\
&	+W \lef   \mathcal{N} \rightarrow \dots, N_{k}+1, N_{k+1}-1, \dots \rig P(\mathcal{N},t) \\
& 	+W \lef \mathcal{N} \rightarrow \dots, N_{k-1}-1,  N_k+1, \dots \rig P(\mathcal{N},t) \\
&	-W \lef \mathcal{N} \rightarrow \dots, N_{k}-1,  N_{k+1}+1, \dots \rig P(\mathcal{N},t) \bigg].
\label{nk_ref2}
\end{aligned}
\end{align}
The transition rates $W$ can now be connected to the reaction rates \refeq{transprob} by considering the proportionality with the number of particles $N_i$ at an individual lattice site,
\begin{equation} 
W\lef \mathcal{N} \rightarrow \dots, N_{i}-1,  N_{i+1}+1, \dots \rig = N_i w_{i\rightarrow i+1} = N_i \Big(1-\frac{N_{i+1}}{N_0}\Big)e^{-E_A+\Delta V_{i}}. 
\end{equation}
With the further introduction of the density of filled oxygen vacancies $n_i := N_i/N_0$,  \refeq{nk_ref2} becomes
\begin{align}
\frac{\mathrm{d}}{\mathrm{d}t} \m{n_k}(t) = 
 - \m{n_k w_{k \rightarrow k-1}} 
+ \m{n_{k+1} w_{k+1 \rightarrow k}}
+ \m{n_{k-1} w_{k-1\rightarrow k}} 
 - \m{n_k w_{k \rightarrow k+1}} 
.
\end{align}
In this part, we introduce the continuum limit of the master equation in a mean field approximation (\textbf{MFA}). 
The MFA now decouples the two point and higher order correlations, resulting in 
\begin{align}
\begin{aligned}
\m{ n_i w_{i\rightarrow j}} =  \m{ n_i } \big( 1-\m{ n_j } \big) e^{-E_A}\m{e^{\Delta V_i}}.
\label{mfa-hop-discrete}
\end{aligned}
\end{align}
Next, we introduce the lattice spacing $\epsilon$ between neighboring sites and define $x :=i \epsilon$. The continuum limit is taken by letting the lattice spacing become infinitesimal while the number of lattice sites $N$ runs to infinity, in such a way that the product of both remains the constant lattice length $\lim_{\epsilon \rightarrow 0, N\rightarrow \infty} \epsilon N = L $. 
Let the averaged density profile be denoted by $\rho(x, t)$, from the relation 
 \begin{equation}
\m{n_i}(t) = \int_{i\epsilon}^{(i+1)\epsilon}\dd x\rho(x,t)
 \end{equation}
 we obtain $\epsilon\rho(x = i\epsilon,t)=\m{n_i}(t)$. The hopping rates that formerly depended upon the density of neighboring lattice sites now depend on those at an infinitesimal distance $\epsilon$, which will obviously introduce an derivation. The continuum counterpart of the voltage drop \refeq{volt-drop-disc} 
 \begin{equation}
\Delta V_i 
\rightarrow \frac{\partial V}{\partial x}\epsilon
\end{equation}
further introduces a sign depending on whether we consider forward ($\epsilon = +|\epsilon|$) or backward ($\epsilon = -|\epsilon|$) fluxes, for it is a directed quantity. Entering this into the MFA hopping rate \refeq{mfa-hop-discrete} yields
\begin{align}
\begin{aligned}
\frac{\partial}{\partial t} \rho(x, t) =& -\sum_{\sigma = \pm 1} \rho(x,t)\big(1-\epsilon\rho(x+\sigma\epsilon,t)\big) e^{-E_A} \exp(+\sigma\partial_x V\epsilon)  \\
 &+\sum_{\sigma = \pm 1} \rho(x+\sigma\epsilon, t)\big(1-\epsilon\rho(x, t)\big) e^{-E_A} \exp(-\sigma\partial_x V\epsilon).
\label{master-cont}
\end{aligned}
\end{align}
The density is expanded up to the second order in $\rho$ according to $\rho(x+\sigma\epsilon, t) = \rho(x,t ) +\sigma \epsilon\partial_x\rho(x,t) + \frac{1}{2}\epsilon^2\partial^2_x\rho(x,t)+ \mathcal{O}(\epsilon^3)$ and the resulting terms are collected with regard to the sign in the exponential, 
%
\begin{align}
	\begin{aligned}
\frac{\partial}{\partial t}  \rho(x,t) =  
&+e^{-E_A}\exp(+\partial_x V\epsilon)\lef	-\epsilon\partial_x\rho + \epsilon^2\lef \frac{1}{2} \partial_x^2\rho + 2 \rho\partial_x\rho\rig \rig \\ 
&+e^{-E_A}\exp(-\partial_x V\epsilon)\lef	+\epsilon\partial_x\rho + \epsilon^2\lef \frac{1}{2} \partial_x^2\rho - 2 \rho\partial_x\rho\rig \rig.
	\end{aligned}
\end{align}
With the definitions of the hyberbolic functions follows
\begin{align}
	\begin{aligned}
\frac{\partial}{\partial t}  \rho(x,t) 
&=	e^{-E_A}\cosh\lef\partial_x V\epsilon\rig 	\epsilon^2\partial^2_x \rho  	
	+ 2e^{-E_A}\sinh\lef\partial_x V\epsilon\rig 	\lef -\epsilon\partial_x\rho +2\epsilon^2\rho\partial_x\rho \rig   \\
&=	e^{-E_A}\cosh\lef\partial_x V\epsilon\rig 	\epsilon^2\partial^2_x \rho		
	-2e^{-E_A}\sinh\lef\partial_x V\epsilon\rig 	\epsilon \partial_x \rho.
\label{protoburgers-hyp}
	\end{aligned}
\end{align}
In the last recast, we have also disregarded the $\epsilon^2$ order in the $\sinh$ term.
As the electrical current through the system is not position dependent, we express local voltage in the following way (cf. \refeq{volt-drop-disc})
\begin{equation}
\partial_x V(x,t) = \frac{V(t)}{\int \dd x' A(x')\rho(x')} A(x)\rho(x) =: I(t) A(x)\rho(x),
\end{equation}
thereby splitting it in a nonlocal and local component. Bearing in mind the continuity equation $\partial_\tau \rho + \mathrm{div} j = 0$, we want to put the spatial derivation in from of the right hand side of \refeq{protoburgers-hyp}. Since the inner derivatives of the hyperbolic functions introduce a further order in $\epsilon$, we disregard these terms and the hyperbolic functions commute with the derivation operator. We gain
\begin{equation}
\frac{\partial}{\partial t}  \rho(x,t) =	\partial_x\big[e^{-E_A}\cosh\lef I A \epsilon \rho\rig 	\epsilon^2\partial_x \rho\big] - \partial_x \big[2e^{-E_A}\sinh\lef I A \epsilon \rho \rig  \epsilon\rho\big] ,
\end{equation}
from which we can easily read the expression for the diffuse and directed oxygen vacancy flux, $j = j_\mathrm{dir} + j_\mathrm{diff}$.  Finally, by further expanding the hyberbolic functions to the first order, we obtain the expression
\begin{equation}
\frac{\partial}{\partial t}  \rho(x,t) =	\partial_x\big[e^{-E_A} \epsilon^2\partial_x \rho\big] - \partial_x \big[2e^{-E_A} I(t) A(x) \epsilon^2\rho^2\big] ,
\label{burgers-appendix}
\end{equation}
By rescaling the time, the prefactor $\epsilon^2 \exp(-E_A)$ can be eliminated. The final equation reads
\begin{equation}
\frac{\partial}{\partial t} \rho(x,t) = \partial_x^2 \rho(x,t) - 2 I(t) \partial_x \big[ A(x)\rho^2(x,t) \big].
\label{burgers-2-appendix}
\end{equation}
It can be compared to the generic viscous Burgers equation, $\partial_t u =  \nu\partial_x^2 u - 1/2 \partial_x u^2 $, with the viscosity $\nu$. So \refeq{burgers-2-appendix} is a Burgers-like equation with a current and resistance profile dependent prefactor.

\section{Notes on the numerics}
\subsection{General remarks}
\label{app-numerics}
Here we specify our assumptions for the form of the resistance profile $A(x)$. While Rozenberg et. al. \cite{Rozenberg2010} used a step profile, for simulation purposes it is feasible to smoothen out the regions between the steps. Also, from a physical point of view, a smoother conduit seems reasonable, since both regions are essentially of the same material. The analytical form was given by \refeq{A_formula}. 
The higher $k$ is, the more rapid the transition from high to low resistivity occurs. For all simulations, we choose the parameter set $k=20$, $L = 1$ and $x_0/L = 0.1$. 
Obviously, the limit $k\rightarrow \infty$ gives a step profile for $A(x)$. For the discretized simulation, the number of lattice sites is set to $N_\text{LS} = 100$.

Given a device of length $L$ that has a certain total resistance $R$, the resistance profile cannot be independent of the number of lattice sites we choose, but scales with it. If for some simulations for example due to numerical expenditure, we need to reduce the number of lattice sites, $A_i$ scales accordingly. In fact, we should think of the resistance profile that is introduced in \refeq{locres} as a resistance per length
\begin{equation}
R = \frac{L}{N_{LS}}\sum\limits_{i} \tilde A_i \frac{N_i}{N_0}, \qquad A_i = \frac{L}{N_{LS}}\tilde A_i.
\end{equation}
We then choose $A_\downarrow$ and $A_\uparrow$ according to the resistances we want to accomplish, and the number of lattice sites, whereas the length $D$ plays no role. 
Concerning the dynamics, the relative differences in $A_i$ are more important than its absolute value, whose influence can just as well be scaled with other parameters, such as the strength of the driving signal, $V_0$. Further calculations are done with $A_\uparrow =100\Omega$ and $A_\downarrow = 1 \Omega$.

The initial distribution of vacancies is chosen in such a way, that all the lattice site contribute equally to the resistance. This choice minimizes the necessary thermalization time of the system. As the individual sites' resistance is proportional to $A_i$ (cf. \refeq{locres}), the individual vacancies are placed at site $i$ according to the probability 
\begin{equation}
P_i = \frac{1}{A_i \sum_i \frac{1}{A_i}}.
\end{equation}
In effect, all sites then contribute with
\begin{equation}
R_0 = N_{LS}\m{R_i} = \frac{N_{LS}}{A_i \sum_i \frac{1}{A_i}},
\end{equation}
for the choice of parameters of the simulations, cf. \fig{dyn-sine}, this yields the normalization $R_0 \simeq 1.76 \Omega$.

\subsection{Numerical integration of the Burgers Equation}

 The generalized Burgers equation is a nonlinear flux transport
 equation, which can be cast into weak form on multiplication by a
 suitable test function, $\psi$ and integrating by parts
 \begin{equation}
 \int_{0}^{1} \frac{\partial \rho}{\partial t}\psi  + \left[\frac{\partial
    \rho}{\partial x} - 2 I(t) A(x)\rho^{2} \right] \frac{\partial
   \psi}{\partial x}\,\mbox{d} x= 0. \label{eq:weak}
 \end{equation}
 The neglected boundary terms ensure that the natural boundary
 conditions are enforced: no flux out of the  domain.

 We discretise solve equation (\ref{eq:weak}) using piecewise quadratic finite
elements for the unknown density $\rho$. The nonlocal term $I(t)$ is handled by treating it as an
independent additional variable with the defining equation
\begin{equation}
 I(t) = \frac{V(t)}{\int_{0}^{1}\mbox{d} x' A(x') \rho(x')}.
\end{equation}
 This equation is assembled in the same
integration loop over the elements used to construct the discrete
nonlinear residuals from equation (\ref{eq:weak}).

 The non-linear residuals are assembled and solved by the C++ library
 \texttt{oomph-lib} \cite{Heil2006a}. Time is advanced by an implicit
 second-order backward difference method (BDF2). The resulting matrix
 system is solved by a direct solver. Adpativity in space and time is
 required to achieve accurate results and it has been confirmed that the
 results do not change if the error tolerances are reduced.

\end{appendix}

\end{document}